\documentclass[tighten, notrackchanges, twocolumn]{aastex62}

\usepackage{amsmath}
\usepackage{amssymb}
\usepackage{mathtools}
\usepackage{mathrsfs}
\usepackage{acronym} 
\usepackage{morefloats}
\usepackage{longtable}
\usepackage{afterpage}
\usepackage{ragged2e}
\usepackage{multirow}
\usepackage[usestackEOL]{stackengine}
\usepackage{hyperref}
\hypersetup{linkcolor=red,citecolor=cyan,urlcolor=magenta}

\usepackage{soul}

\usepackage[makeroom]{cancel}
\usepackage{xcolor}

\graphicspath{{figs/}}

\usepackage{xspace}

\newcommand{\msol}{M$_\odot$\xspace}
\newcommand{\joo}{PSR~J0030$+$0451\xspace}
\newcommand{\jos}{PSR~J0740$+$6620\xspace}
\newcommand{\jof}{PSR~J0437$-$4715\xspace}
\newcommand{\sch}{Schwarzschild\xspace}

\defcitealias{Bogdanov19b}{B19}
\defcitealias{Bogdanov21}{B21}

\shorttitle{Exploring Waveform Variations among NS Ray-tracing Codes}
\shortauthors{Choudhury~et~al.}

\begin{document}

\title{Exploring Waveform Variations among Neutron Star Ray-tracing Codes for Complex Emission Geometries}

\correspondingauthor{D.~Choudhury}
\email{D.Choudhury@uva.nl}

\author[0000-0002-2651-5286]{Devarshi~Choudhury}
\affil{Anton Pannekoek Institute for Astronomy, University of Amsterdam, Science Park 904, 1090GE Amsterdam, the Netherlands}

\author[0000-0002-1009-2354]{Anna~L.~Watts}
\affil{Anton Pannekoek Institute for Astronomy, University of Amsterdam, Science Park 904, 1090GE Amsterdam, the Netherlands}

\author[0000-0001-6157-6722]{Alexander~J.~Dittmann}
\affil{Department of Astronomy and Joint Space-Science Institute, University of Maryland, College Park, MD 20742-2421, USA}

\author[0000-0002-2666-728X]{M.~Coleman Miller}
\affil{Department of Astronomy and Joint Space-Science Institute, University of Maryland, College Park, MD 20742-2421, USA}

\author[0000-0003-4357-0575]{Sharon M.~Morsink}
\affil{Department of Physics, University of Alberta, Edmonton, AB T6G 2G7, Canada}

\author[0000-0001-6356-125X]{Tuomo~Salmi}
\affil{Anton Pannekoek Institute for Astronomy, University of Amsterdam, Science Park 904, 1090GE Amsterdam, the Netherlands}

\author[0000-0003-3068-6974]{Serena~Vinciguerra}
\affil{Anton Pannekoek Institute for Astronomy, University of Amsterdam, Science Park 904, 1090GE Amsterdam, the Netherlands}

\author[0000-0002-9870-2742]{Slavko~Bogdanov}
\affil{Columbia Astrophysics Laboratory, Columbia University, 550 West 120th Street, New York, NY 10027, USA}

\author[0000-0002-6449-106X]{Sebastien~Guillot}
\affil{IRAP, CNRS, 9 avenue du Colonel Roche, BP 44346, F-31028 Toulouse Cedex 4, France}
\affil{Universit\'{e} de Toulouse, CNES, UPS-OMP, F-31028 Toulouse, France.}

\author[0000-0002-4013-5650]{Michael~T.~Wolff}
\affil{Space Science Division, U.S. Naval Research Laboratory, Washington, DC 20375, USA}

\author[0009-0008-6187-8753]{Zaven~Arzoumanian}
\affil{X-Ray Astrophysics Laboratory, NASA Goddard Space Flight Center, Greenbelt, MD 20771, USA}

\begin{abstract}


Pulse Profile Modeling (PPM), the technique used to infer mass, radius and geometric parameters for rotation-powered millisecond pulsars using data from the Neutron Star Interior Composition Explorer (NICER), relies on relativistic ray-tracing of thermal X-ray photons from hot spots on the neutron star surface to the observer.  To verify our ray-tracing codes we have in the past conducted cross-tests for simple hot spot geometries, focusing primarily on the implementation of the space-time model. In this paper, we present verification for test problems that explore the more complex hot spot geometries that are now being employed in the NICER PPM analyses. We conclude that the accuracy of our computed waveforms is in general sufficiently high for analyses of current NICER data sets.  We have however identified some extreme configurations where extra care may be needed.
\end{abstract}

\section{Introduction}
\label{sec:intro}

Pulse Profile Modeling (PPM) of X-ray spectral-timing data from Rotation-powered Millisecond Pulsars (MSPs), enabled by NASA's {\it Neutron Star Interior Composition Explorer} (NICER), is now an established technique for measuring neutron star mass and radius \citep{Riley19,Miller19,Riley21,Miller21,Salmi22,Salmi23,Afle23,Vinciguerra23,Vinciguerra24,Salmi24,Dittmann24,Choudhury24,Miller24}. By measuring masses and radii, it is possible to constrain the Equation of State (EoS) of the cold ultradense matter in the neutron star core \citep[see e.g.][]{Miller21,Raaijmakers21,Legred21,Biswas22,Huth22,Takatsy23,Annala23,Rutherford24}. PPM also allows one to map the hot X-ray emitting regions on the pulsar's surface (thought to originate from the magnetic poles), shedding light on both pulsar emission mechanisms and magnetic field structure and evolution
\citep[see e.g.][]{Bilous19,Chen20,Kalapotharakos21,Das22,Carrasco23}. 

PPM involves relativistic ray-tracing of thermal emission from the stellar surface to the observer, exploiting a computationally-efficient approximation for the space-time of rapidly rotating neutron stars \citep{Miller98,Poutanen03,Morsink07} and a suite of appropriate atmosphere models \citep{Ho01}. The data that we are modeling, the {\it pulse profile}, is a rotational phase-resolved count spectrum built up over very long exposure times on the order of megaseconds, see Figure 3 of \citet{Bogdanov19a} for examples. Full details of the ray-tracing computation and the Bayesian inference process used in the NICER PPM analysis can be found in \citet{Bogdanov19b,Bogdanov21}, hereafter \citetalias{Bogdanov19b,Bogdanov21}.  

In order to have  confidence in the results of our PPM analysis, it is important to verify our ray-tracing codes. This means checking their accuracy and precision over the range of conditions that we expect to encounter in our analyses, including (but not limited to) different masses, radii, spin rates, observer inclinations, and hot spot properties.

\citetalias{Bogdanov19b} conducted a first set of ray-tracing cross-tests on a suite of test problems, assuming a single uniform temperature circular hot spot and blackbody emission. Factors that were tested included the space-time approximation, spot size and location, observer inclination and stellar spin rate. Simulated pulse profiles from several different codes were compared, and checked against the results from exact numerical space-time calculations.  \citetalias{Bogdanov21} carried out additional cross-tests to verify the implementation of multiple-imaging\footnote{Highly curved photon trajectories can lead to photons orbiting the star before arriving to the observer, resulting in multiple imaging of the star. Primary images imply deflection angles $< \pi$ radians, secondary images consist of deflection angles between $\pi$ and $2\pi$ radians, and so on.}, that become especially important for highly compact neutron stars. The multiple-imaging cross-tests comprised blackbody radiation\footnote{Note that the waveforms used in the parameter recovery tests reported in \citetalias{Bogdanov21} did implement atmosphere and ISM effects, but agreement between the different codes at this stage was not reported.} from a globally isotropic surface radiation field (for zero rotation) or two uniform-temperature circular hot spots (finite rotation).  The agreement between codes reported in these papers was at a level where expected statistical errors accompanying observations should dominate over any systematics arising from approximations used in the ray-tracing implementations. 

In practice, the various ray-tracing codes being used for PPM within the NICER collaboration admit more complex patterns than uniform-temperature circular hot spots.  
Theoretical models of the magnetic polar caps of MSPs hint at surface heating distributions that cannot simply be encapsulated by circular hot spots \citep[see e.g.,][]{Harding11, Timokhin13, Gralla17, Lockhart19}.  Moreover, results from the NICER PPM analysis of \joo and \jof indicate that rotation-powered MSPs can indeed have more complex and elongated surface emission patterns \citep{Riley19, Miller19, Vinciguerra24,Choudhury24,Miller24}. More complex hot spot shapes have also been explored for \jos, although at present they do not seem to be necessary to explain the NICER data \citep{Riley21,Miller21} and are disfavored by XMM-Newton data \citep{Dittmann24}.  

Pulse profiles generated by these more complex hot spot geometries may be more sensitive to the exact treatments adopted by the different ray-tracing codes. Particular aspects of concern include hot spots formed from overlapping emitting elements, elongated shapes, and multiple-imaging for highly compact systems. In this paper, we report cross-tests for a suite of test problems designed to explore these aspects. 

\section{Ray-tracing codes}
\label{sec:ray-tracing}

There are three different relativistic ray-tracing codes involved in the suite of tests described in this paper: IM, Alberta, and X-PSI. The Illinois-Maryland \citep[IM, ][]{Miller98,Lamb09,Lo13,Miller15}, Alberta \citep{Cadeau07,Morsink07,Stevens16}, and X-ray Pulse Simulation and Inference \citep[X-PSI, ][]{Riley23}\footnote{\url{https://github.com/xpsi-group/xpsi}} codes are described in detail in Appendices B.3, B.4 and B.5 of \citetalias{Bogdanov19b} respectively\footnote{Note that X-PSI was then referred to as the AMS code.}.  All three codes were involved in the cross-tests described in \citetalias{Bogdanov19b}\footnote{The full suite of tests in that paper considered both the \sch + Doppler (S+D) \citep{Miller98,Poutanen03} and the Oblate \sch (OS) space-time approximations \citep{Morsink07}, the OS now being the one in use for NICER analysis. Of the three ray-tracing codes being compared in this paper, the S+D tests were conducted between the Alberta and IM codes, while the OS tests involved all three.}. The multiple-imaging tests for very compact stars reported in \citetalias{Bogdanov21} involved the IM and X-PSI codes (the Alberta code does not have multiple-imaging capabilities). 
All NICER MSP PPM analyses published to date (in which the ray-tracing simulation software is coupled to statistical sampling software to carry out parameter inference) have used the IM code \citep{Miller19,Miller21,Dittmann24,Miller24} and the X-PSI code \citep{Riley19,Riley21,Salmi22,Salmi23,Vinciguerra24,Salmi24,Choudhury24}. 

As described in Section \ref{sec:intro}, theoretical models of the heated magnetic polar caps of MSPs predict a variety of complex shapes.  The model spaces of the three codes being compared in this paper are designed to capture such complexities in a rudimentary manner, utilizing overlapping circles (X-PSI and Alberta codes) or ellipses (IM code) on the neutron star surface. The tests reported in \citetalias{Bogdanov19b} and \citetalias{Bogdanov21} used one or two non-overlapping uniform-temperature circular hot spots.  However as mentioned in the previous section, ongoing NICER analysis also involves more complex hot spot geometries. 

The pulse profiles generated by such complex hot spot geometries could potentially be more sensitive to the exact treatments adopted by the different ray-tracing codes. It is therefore important to expand the set of tests conducted by \citetalias{Bogdanov19b} and \citetalias{Bogdanov21} to ensure that they yield consistent results for various cases, assuming the same input. The ability to generate near-identical pulse profiles is key to eliminating one source of potential discrepancies in the follow-up parameter estimation of real data. 

We have identified the following areas that require further testing in this context: 
\begin{itemize}
    \item Spot overlap treatment -- Overlapping spots can be used to approximate temperature gradients, and annular, crescent-like, or elliptical shapes. When individual circles overlap, an order of precedence needs to be established as to whose surface area is to be considered for evaluation of local radiation intensities. The handling of such overlaps varies between the codes, with X-PSI especially differing in its method of surface discretisation (see appendix B.5 of \citetalias{Bogdanov19b}). 
    \item Elongated shapes -- Since the boundaries of the circular spots are discretised in some fashion in all three codes, edge effects of overlapping spots can become especially important when dealing with elongated structures such as a large and/or thin (possibly non-concentric) annulus or crescent.
    \item Multiple-imaging for compact stars -- In the case of highly compact stars such as those tested in \citetalias{Bogdanov21} using simple circular hot spots, the flux received by a distant observer travels along multiple different paths. Such tests are extended to more complex spot geometries.
\end{itemize}

In the following section, we elaborate on the models selected to test the aforementioned situations and compare the corresponding code outputs.  As in \citetalias{Bogdanov19b} and \citetalias{Bogdanov21}, we will, in general, illustrate our results by showing the energy-summed waveform, which in this paper we refer to as the {\it waveform}\footnote{In previous papers the terms pulse profile and waveform have often been used interchangeably.}.  However, in all cases we also compare the full energy-resolved pulse profiles for discrepancies.  One key difference compared to \citetalias{Bogdanov19b} and \citetalias{Bogdanov21} is that in those papers, the cross-comparisons were done before applying the instrument response (that captures the instrument effective area and the mapping of  incident photons of a given energy to specific detector energy channels) and correcting for interstellar absorption; in this paper comparisons are done afterwards so that we can also check these aspects.

\section{Waveform and Pulse Profile Comparisons}
\label{sec:waveform comparisons}

We have developed five extreme test-cases to probe the resultant differences of the varying treatments in the three ray-tracing codes, in the context of the situations listed in the previous section. Four of these tests involve only a single-temperature hot spot and the fifth test involves a two-temperature hot spot.  In order to create complex shapes, we then introduce a \textit{masking} component atop the emitting single-temperature component that blocks out any emission originating from within that region. The two-temperature spot involves two overlapping emitting components, of which one is a \textit{ceding} component, i.e., emission from the overlapped subset of this component is discounted. Each individual component in these tests, be it emitting or masking, are circular since all three ray-tracing codes can accommodate such configurations.

All the tests share the following non-spot model parameters:
\begin{itemize}
    \item Distance = 150 pc.
    \item Observer inclination angle = 2.391 radians (with respect to rotational North Pole).
    \item Spin frequency (as seen by a distant observer) = 300 Hz.
    \item Neutral Hydrogen column density (N$_{\rm H}$) = $0.2 \times 10^{20} {\rm cm}^{-2}$. 
\end{itemize}

The inclination angle is chosen to agree with the value measured for PSR~J0437--4715 \citep{Reardon24} and results in a viewing angle where the pole closer to the observer is always visible and the opposite pole is occulted (except for the model with very high compactness, described below, where the opposite pole is also visible due to strong gravitational lensing). The choice of spin frequency is motivated by the fastest NICER rotation-powered MSP target, \jos, which has a frequency of 346 Hz. \citetalias{Bogdanov19b} also showed that OS waveforms match the corresponding precise numerical waveforms within an accuracy of $<0.1\%$ for spin frequencies below 300 Hz. The discrepancies increase upon moving to much higher spin frequencies, above a few percent for 600 Hz and higher \citep{Cadeau07, Pihajoki2018}.

All of the stellar models that consist of a single-temperature hot spot configuration have a gravitational mass of 1.4 \msol and an equatorial circumferential radius of 12.0 km. 
The two-temperature spot model has a gravitational mass of 2.14 \msol and an equatorial circumferential radius of 9.6 km, following the choice of \citetalias{Bogdanov21}, to create a highly compact star. 

\begin{figure*}[h!]
    \centering
    \includegraphics[width=0.85\linewidth]{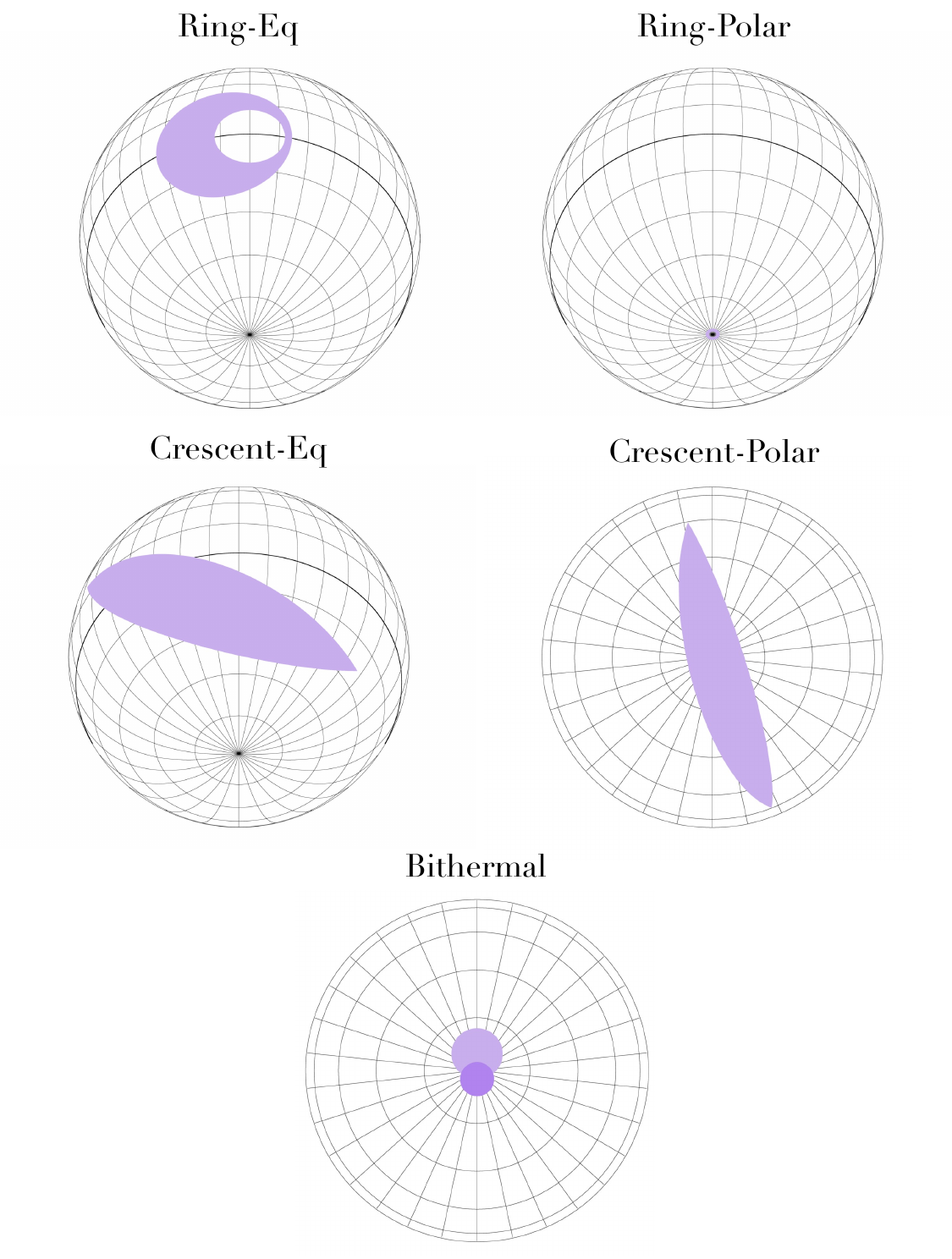}
    \caption{Resultant hot spot shapes based on the geometry parameters prescribed in Table \ref{tab:geometries}. Full animation sets showing the spot geometries as the star rotates are available in Zenodo. The purple shaded regions represent the X-ray emitting areas. The Ring-Eq, Ring-Polar and Crescent-Eq spots are viewed from the observer inclination angle of 2.391 radians and include the effects of gravitational lensing. The darker latitude represents the equator for the three aforementioned configurations. The Crescent-Polar and Bithermal spots are viewed from the North Pole and do not include lensing effects for visualization purposes. The lighter shaded region of the Bithermal spot represents the ceding emission component. Doppler aberrations are not shown for any of the plots to make it easier to discern the spot shapes.}
    \label{fig:geometries}
\end{figure*}

Two of the single-temperature models consist of nested non-concentric masking components. The consequent non-concentric annular emission geometries, simply referred to as \textit{rings} hereon, are meant to test the spot overlap treatment between the codes. We also vary their sizes to probe the effects of spot discretisation. The other two single-temperature models consist of protruding masking elements, yielding crescent-shaped emission geometries, simply referred to as \textit{crescents} hereon. The crescents are parameterized such that they are highly elongated, thus emphasizing edge effects. The components of the two-temperature spots are not nested, i.e. the ceding emission component protrudes out of the main emission component. This geometric configuration is simply referred to as the \textit{bithermal} spot hereon. The bithermal spot is placed on the North Pole. Given the viewing angle, this spot would have been permanently occulted in the absence of light bending.  The high compactness of the star results in the spot always being visible.  Such a configuration also results in multiple-imaging. However, atmospheric beaming effects reduce the number of surface tangential photon paths taken which affects the consequent pulse profile. All hot spot parameters pertaining to these tests are detailed in Table \ref{tab:geometries}, and the corresponding shapes are shown in Figure \ref{fig:geometries}.

\begin{table*}[t!]
\renewcommand{\arraystretch}{1.5}
    \centering
    \caption{hot spot parameters for the waveform tests.}\begin{tabular}{l c c c c c} \hline \hline
        \textbf{Geometry parameters}& \textbf{Ring-Eq}& 
        \textbf{Ring-Polar} & 
        \textbf{Crescent-Eq}& 
        \textbf{Crescent-Polar}& 
        \textbf{Bithermal}\\ \hline
                 \textbf{Emitting component} & & & & & 
                 \\
         Colatitude $\theta_{e}$ (rad)& $\dfrac{\pi}{2}$& $\pi - 0.001$& 1.0& 1.15&0.05\\
         Azimuth $\phi_{e}$ (rad)& $-0.2$ & $-0.02$ & $-1.0$& $-0.1$&0.0\\
         Angular radius $\zeta_{e}$ (rad)& 0.5& 0.05& 1.4& 1.3&0.1\\
         Temperature (log$_{10}$T[K])& 6.0& 6.0& 6.0& 6.0&6.2\\
         \hline
                \textbf{Masking/ceding component} & & & & & \\
         Colatitude $\theta_{m/c}$ (rad)& $\dfrac{\pi}{2}$& $\pi - 0.001$& 0.5& 1.75&0.1\\
         Azimuth $\phi_{m/c}$ (rad)& 0.0& 0.0& 0.0& 0.0&$\pi$\\
         Angular radius  $\zeta_{m/c}$ (rad)& 0.25& 0.025& $\dfrac{\pi}{2} - 0.001$& $\dfrac{\pi}{2} - 0.001$&0.15\\ 
         [1.0ex] \hline
    \end{tabular}
    \tablecomments{The non-overlapped subset of the ceding component emits X-rays with an effective temperature of $10^{5.8}$ K. The suffixes Eq and Polar denote the (unmasked) emitting regions passing over the equator and encompassing one of the rotational poles respectively. All configurations assume an observer inclination angle of 2.391 radians. All of the single temperature spot configurations are for a 1.4 \msol and 12.0 km radius neutron star, and the bithermal configuration is for a 2.14 \msol and 9.6~km neutron star.}
    \label{tab:geometries}
\end{table*}

\begin{table}
    \centering
    \caption{X-PSI resolution setting classifications}\begin{tabular}{l c c c c} \hline \hline
        \textbf{X-PSI setting}& \textbf{Low}& \textbf{Std} & \textbf{High}& \textbf{Ultra}\\ \hline
         \texttt{sqrt\_num\_cells}& 16& 32& 64& 256\\
 \texttt{min\_sqrt\_num\_cells}& 16& 32& 64&256\\
 \texttt{max\_sqrt\_num\_cells}& 16& 32& 64&256\\
         \texttt{num\_leaves}& 32 & 64& 128& 512\\
         \texttt{num\_energies}& 64& 128& 256& 512\\
         \texttt{num\_rays}& 100& 512& 640& 1024\\
         \hline
    \end{tabular}
    \tablecomments{For details on the exact definitions and application of these resolution settings, we refer the reader to the X-PSI documentation webpage\footref{footnote: xpsi docs}.}
    \label{tab:resolutions}
\end{table}

For all models, we define the hot spot emitting components as having a geometrically thin, fully-ionized hydrogen atmosphere, with corresponding specific intensities determined from \texttt{NSX}-generated lookup tables \citep{Ho01}. All three codes first compute the signal for a discrete set of photon energies (in the observer's reference frame) which are then attenuated by the appropriate factor for the value of N$_{\rm H}$ specified above.  

The set of energies over which the initial signals are computed varies between the different codes. X-PSI uses a log-spaced energy grid, with more points towards the lower energies where more emission is expected. The grid density is user-defined via the \texttt{num\_energies} setting (see table \ref{tab:resolutions} for settings used in this work). The IM code uses a linearly spaced initial coarse energy grid (with a spacing of 0.1 keV), which is then quadratically interpolated to a spacing of 0.1 eV. The Alberta code uses a linearly spaced energy grid with 5 eV intervals, corresponding directly to the central value for each NICER energy channel.

Each of the three codes then use the same interstellar absorption table to produce the absorbed emission. To obtain the interstellar attenuation factor as a function of energy,  we use tables generated using the \texttt{tbnew}\footnote{\url{https://pulsar.sternwarte.uni-erlangen.de/wilms/research/tbabs/}} model \citep[][updated in 2016]{Wilms00}. This emission is finally convolved with the NICER XTI response matrix, via matrix multiplication, using the 3C50 response files used in \citet{Choudhury24, Miller24} to obtain the instrument registered counts. The pulse profiles are generated over 32 phase bins and instrument channel range [30, 300)corresponding to photon energies in the range 0.3 - 3.0 keV. We specify an exposure time of 1~Ms for each test case. The interpolation routines vary between the three codes when mapping the energies from the attenuation table onto the respective energy grids, and when obtaining the binned emission corresponding to the NICER energy intervals.

For the X-PSI waveforms, we also test the effects of different resolution settings\footnote{\label{footnote: xpsi docs}\url{https://xpsi-group.github.io/xpsi/hotregion.html}}, listed in table \ref{tab:resolutions}. Our \textit{Low} resolution settings used in this paper are chosen such that they are either equal to or lower than the low-resolution settings used in \citet{Vinciguerra23, Vinciguerra24, Choudhury24}\footnote{The low-resolution settings in these other works have a higher value of \texttt{sqrt\_num\_cells}=18 and \texttt{num\_rays}=512.}. The standard (\textit{Std})  resolution settings are set around the standard values used in the different works using X-PSI. The \textit{High} resolution settings are chosen to roughly double the standard settings while maintaining computational tractability in the context of inference runs. The \textit{Ultra} resolution settings are chosen to be arbitrarily high to benchmark the other settings (and other codes) and test resolution setting convergence. The corresponding waveform calculation using the \textit{Ultra} settings are far too computationally expensive for any inference run. The approximate computation time for generating a pulse profile by each of the codes is shown in table \ref{tab:comp_time}.

\begin{table}
    \centering
    \caption{Approximate pulse profile computation time}\begin{tabular}{l c} \hline \hline
        \textbf{Code}& \textbf{Computation time}\\ \hline
         X-PSI low& $\lesssim 0.1$ s\\
         X-PSI std& $\sim 0.1 - 0.5$ s\\
         X-PSI high& $\sim 0.3 - 5$ s\\
         X-PSI ultra& $\sim 1-5$ min\\
         IM& $\sim 5-10$ s\\
         Alberta& $\sim 5-30$ s\\
         \hline
    \end{tabular}
    \tablecomments{The exact times depend on the model chosen and the CPU architecture of the system the code is being run on, and therefore we only provide approximate times.}
    \label{tab:comp_time}
\end{table}

In the following subsections, we showcase the results of each of the ray-tracing codes for the various test cases. The pulse profiles generated by the different groups and a Jupyter notebook to reproduce the comparison plots shown in the paper are provided in a Zenodo\footnote{\url{https://doi.org/10.5281/zenodo.13133749}} repository. X-PSI being an open-source software, we have also provided the relevant modules for generating the pulse profiles using X-PSI.

\begin{figure*}
    \centering
    \includegraphics[width=0.9\linewidth]{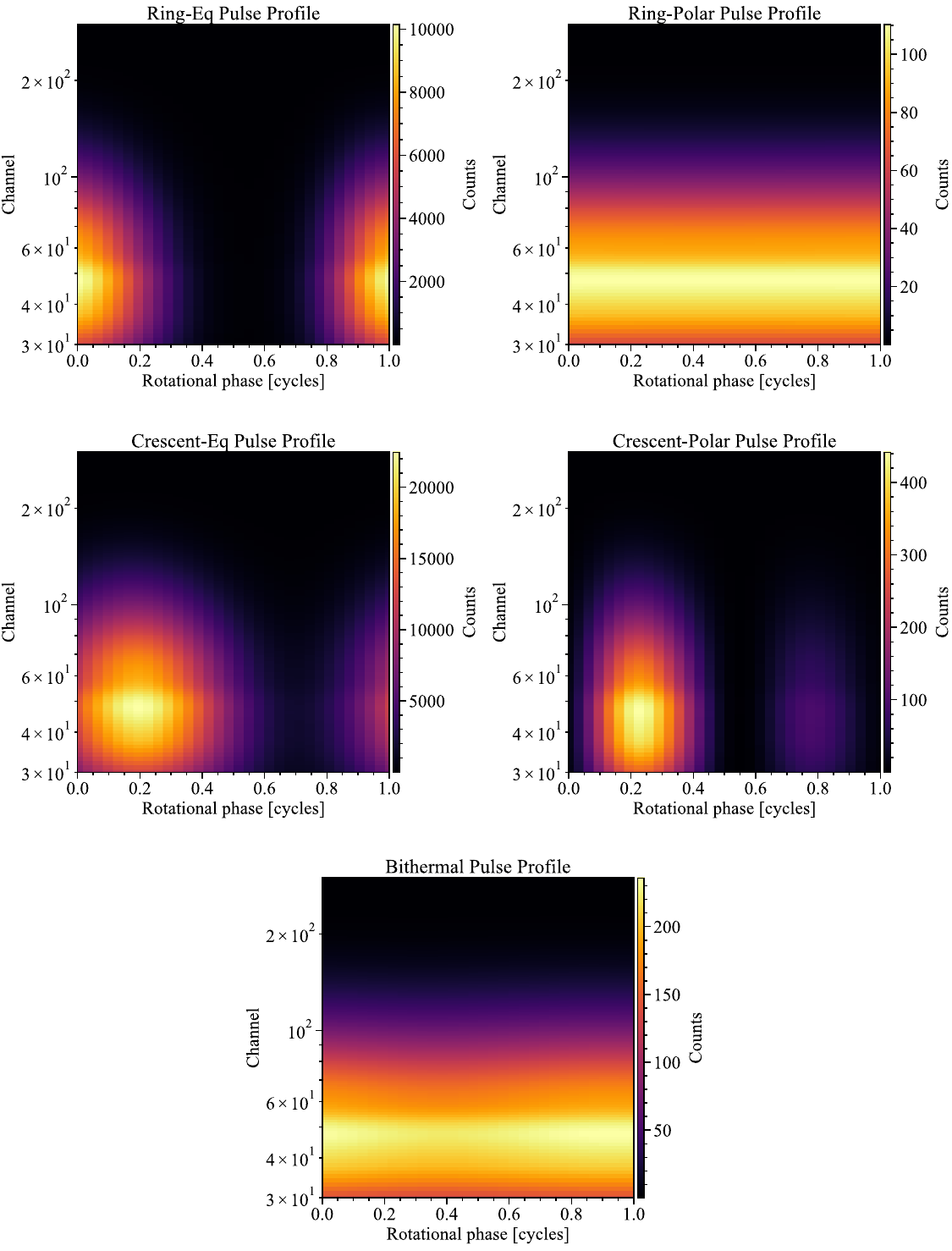}
    \caption{Simulated phase-energy resolved X-ray pulse profiles, over 32 phase bins and instrument channel range [30,300), generated by the different test configurations, as registered by NICER, using the 3C50 instrument response. These pulse profiles are generated by X-PSI using the ultra resolution settings.}
    \label{fig:pulse profiles}
\end{figure*}

\subsection{Test: Ring-Eq spot}
\label{sec:est-eq}

The Ring-Eq configuration yields a pulse profile (Figure \ref{fig:pulse profiles}) consisting of a singular prominent pulse per cycle and reaching a high number of counts ($\sim10000$) in a phase-energy bin at its peak. 

\begin{figure}
    \centering
    \includegraphics[width=1\linewidth]{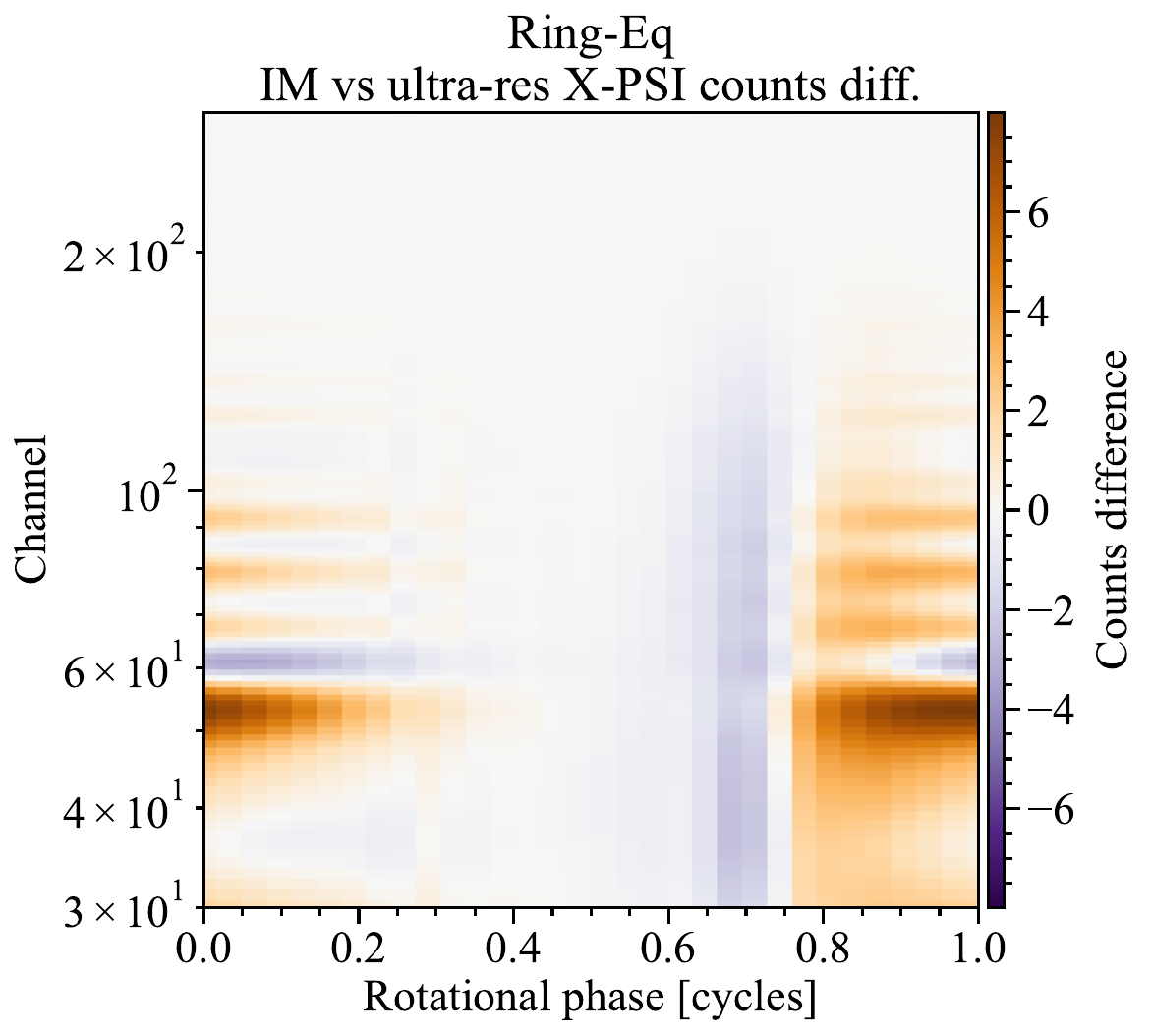}
    \caption{Total counts difference between the full phase-energy-resolved pulse profiles generated by the IM code and X-PSI (using ultra resolution settings).}
    \label{fig:est-eq pp diff ultra-xpsi and IM}
\end{figure}

To compare the different code outputs and X-PSI resolution settings, we checked the total and fractional differences in the energy-summed waveforms, the phase-summed spectra, and the full phase-energy-resolved pulse profile. The last check mentioned is to search for any correlated residual structures upon comparing any two pulse profiles that are not apparent in the former checks. We found no such specific correlated differences for any of the test cases that cannot be explained by looking at the energy-summed waveforms and the phase-summed spectra, and therefore refrain from showing them beyond the Ring-Eq case as an example in Figure \ref{fig:est-eq pp diff ultra-xpsi and IM} (The full set of phase-energy-resolved residuals are available in Zenodo). \textbf{Note that when comparing the waveforms and spectra, we always refer to the fractional difference as compared to the ultra resolution X-PSI waveform, unless explicitly stated otherwise.} 

\begin{figure*}
    \centering
    \includegraphics[width=1\linewidth]{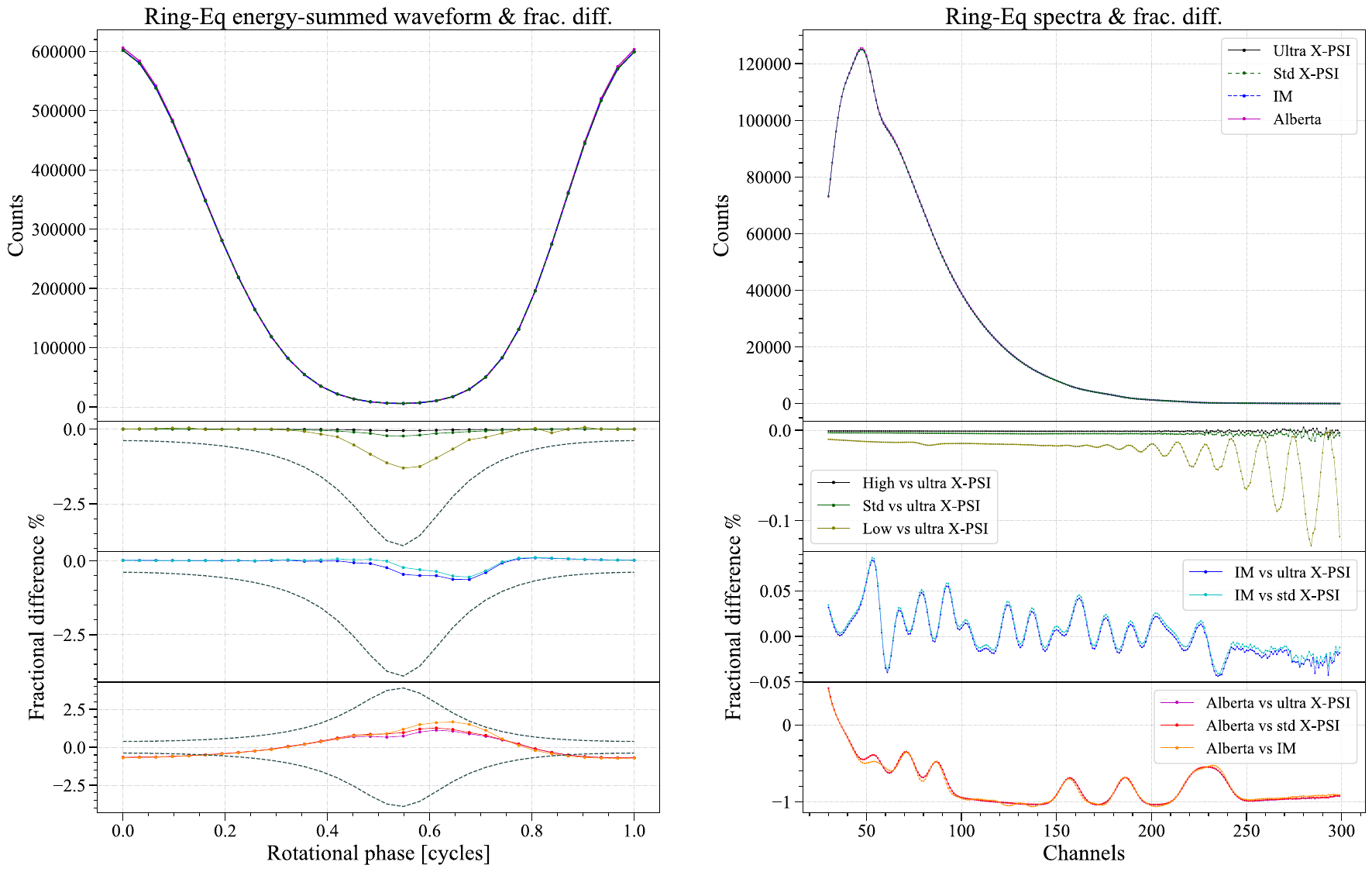}
    \caption{Energy-summed waveforms (left) and phase-summed spectra (right) generated by the different codes for the Ring-Eq configuration (top), the corresponding fractional differences between the different X-PSI resolution settings as benchmarked against the ultra settings (top-middle), the fractional differences between the IM code against the X-PSI standard and ultra resolution settings (bottom-middle), and the fractional differences between the Alberta code against the IM code, the X-PSI standard and ultra resolution settings (bottom). The grey dashed lines in the fractional difference plots indicate the expected Poisson fluctuation for the X-PSI ultra waveform scaled to yield a total of $10^6$ counts. The legends shown in the spectral plot also apply to the waveform plot.}
    \label{fig:est-eq frac diffs}
\end{figure*}

Upon inspecting the waveforms, we find that all the X-PSI resolution settings yield relatively consistent results, with slightly larger variations for the low resolution waveform. The fractional differences (Figure \ref{fig:est-eq frac diffs}) are very small ($<0.1\%$) except towards the waveform minima, where the model counts drop by nearly two orders of magnitude compared to the maxima. The maximum fractional difference for the low resolution waveform is $1.3\%$. The fewer spot cell meshes prescribed by the low resolution settings implies that larger emitting areas abruptly go out of view as the star rotates, instead of a smooth transition, leading to an under-prediction of counts.

The IM waveform also exhibits small differences, generally below $0.1\%$, reaching a maximum fractional difference of $0.63\%$ towards the waveform minima, at slightly later phases compared to the different X-PSI waveforms. The Alberta waveform seems to deviate more from the other waveforms across all phases, including at the waveform maxima. It exhibits a maximum fractional difference of $1.13\%$ against X-PSI ultra, and $1.67\%$ against the IM waveform.

 Figure \ref{fig:est-eq frac diffs} also shows the expected Poisson noise level per phase bin corresponding to a scaled X-PSI ultra waveform, where the waveform is normalized such that the total number of counts equals $10^{6}$ (also done for all other figures showing Poisson bands), as is typically the case in the available NICER datasets. The IM and X-PSI discrepancies per phase bin are well below the expected Poisson fluctuations and can therefore be safely subsumed by the noise. The deviations in the Alberta waveforms are comparable to Poisson fluctuations in the phase bins with the highest counts.

Comparing the spectral differences between the different X-PSI settings, again we find generally consistent results. The lower the resolution, the fewer the registered counts across nearly all channels. This is simply a consequence of the under-prediction of counts for the waveform as the spot goes out of view. The low resolution spectra also exhibit oscillations in the fractional differences towards higher channels, the amplitude of which grows as the counts tend to zero (see top-middle panel of the plot on the right in Figure \ref{fig:est-eq frac diffs}). This is likely a consequence of using lower energy resolution when generating the photospheric emission and then interpolating onto the instrument response energy grid. 

The fractional differences between IM and X-PSI spectra also exhibit oscillatory behavior but over all energy channels and with varying amplitudes. Varying oscillations are also found for the Alberta spectrum, although with much larger amplitudes. These are likely consequences of different model energy grid definitions and interpolation schemes opted by the different codes. Although not mentioned explicitly in that paper, such waviness was also present in the tests conducted in \citetalias{Bogdanov19b}.

\subsection{Test: Ring-Polar spot}
\label{sec:est-polar}

\begin{figure*}
    \centering
    \includegraphics[width=1\linewidth]{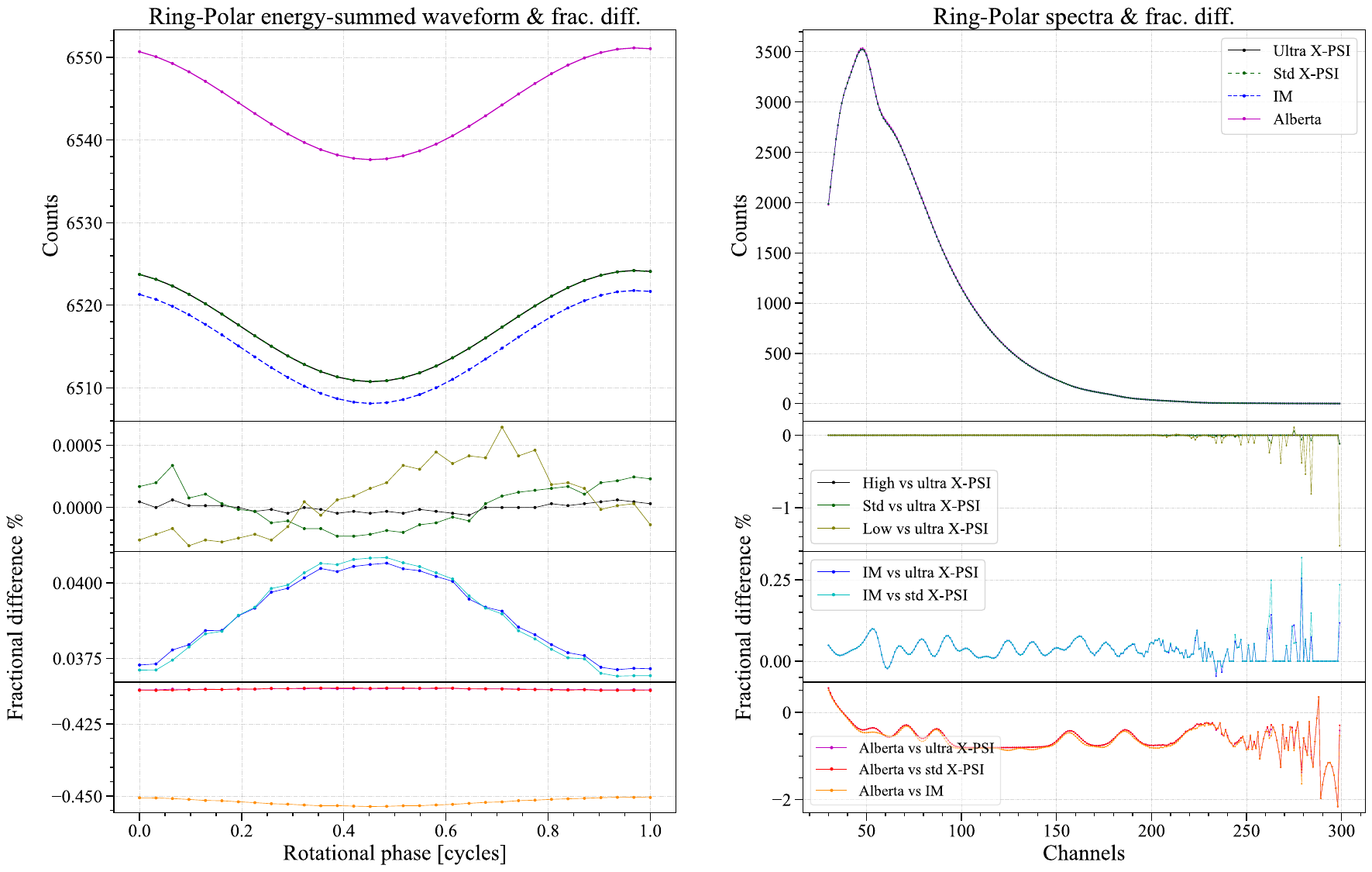}
    \caption{Energy-summed waveforms (left) and phase-summed spectra (right) generated by the different codes for the Ring-Polar configuration (top) and the corresponding fractional differences. For clarity, the Poisson bands are not shown in this figure since the fractional differences are far lower than the noise level.}
    \label{fig:est-polar frac diffs}
\end{figure*}

The pulse profile generated by the Ring-Polar configuration (Figure \ref{fig:pulse profiles}) does not exhibit prominent pulse modulation since the hot spot is placed at the pole, and due to its small size, registers low number of counts ($\sim 100$ counts in the brightest phase-energy channels). However, there are very slight pulsations present, as can be seen in the waveform in Figure \ref{fig:est-polar frac diffs}, caused by the off-centered masking region. 

The fractional differences between the different codes and X-PSI resolution settings are generally at a much lower level than for the Ring-Eq waveform. The apparent dissimilarities in the Ring-Polar waveforms predicted by the different codes actually consist of only small absolute count differences, and are a consequence of hot spot discretisation sensitivities for a tiny emitting region that generates an overall low number of counts. 

In this case, we find more erratic patterns of fluctuations in the waveforms as we vary the X-PSI resolution settings, with no apparent trends as we increase or decrease resolution (see top-middle panel of the plot on the left in Figure \ref{fig:est-polar frac diffs}). All resolution settings are highly consistent with each other with minuscule fractional differences with respect to the X-PSI ultra setting. The maximum fractional difference exhibited by the low resolution setting is $0.0006\%$.

The IM waveform shows larger differences compared to the X-PSI waveforms, than between the different X-PSI resolutions, although these are also very minor, peaking towards the waveform minima with a maximum fractional difference of $0.04\%$. The Alberta waveform also shows rather small differences compared to the other codes in this case, although the fractional differences are an order of magnitude higher than the differences shown by IM waveforms. The Alberta waveform has a maximum fractional difference of $0.41\%$ against the X-PSI ultra waveform and $0.45\%$ against the IM waveform.

All these differences are well below the expected Poisson noise, where the minimum noise level at the highest count phase bin is $1.64\%$ (assuming X-PSI ultra waveform scaled to $10^6$ counts), and therefore we do not show the Poisson band in the figure.

The spectral differences between the different X-PSI settings are also lower for this model. The oscillatory behavior in fractional difference seen for Ring-Eq by the low resolution setting is not as prominent and affects fewer of the upper channels, leading to the more jagged appearance. 

The energy interpolation related oscillations in fractional difference seen for the IM and Alberta spectra seem to maintain a similar pattern as seen previously, except at the higher channels due to fewer counts being present, which increases the channel-to-channel variation.

\subsection{Test: Crescent-Eq spot}
\label{sec:pst-eq}

\begin{figure*}
    \centering
    \includegraphics[width=1\linewidth]{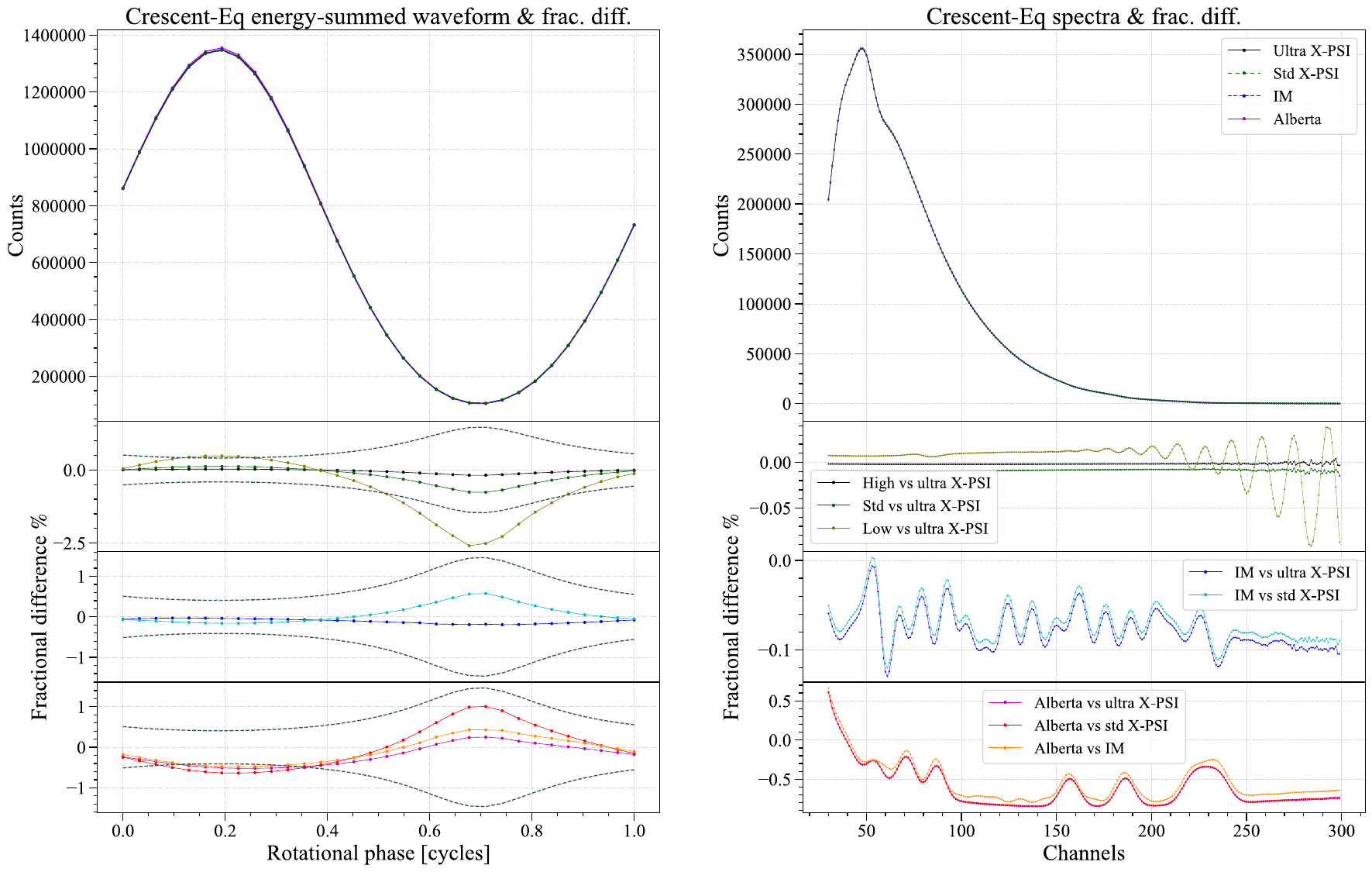}
    \caption{Energy-summed waveforms (left) and phase-summed spectra (right) generated by the different codes for the Crescent-Eq configuration (top) and the corresponding fractional differences.}
    \label{fig:pst-eq frac diffs}
\end{figure*}

The Crescent-Eq configuration results in a large elongated spot that is easily visible over many phases, leading to a very high number of counts being registered ($\sim 22000$ counts in the brightest phase-energy bins), and thus yields a pulse profile consisting of a wide and prominent singular pulse (Figure \ref{fig:pulse profiles}). This Crescent-Eq waveform (Figure \ref{fig:pst-eq frac diffs}) appears to be slightly more sensitive to the different code treatments and resolution settings than seen previously, especially at both the waveform crests and troughs (the latter more than the former). 

The high resolution X-PSI waveform nominally varies with respect to the ultra settings, exhibiting fractional differences below $0.1\%$ across most phases, and a maximum fractional difference of $0.18\%$ at the waveform trough. This is comparable to the level of variation between the IM and X-PSI ultra waveforms, which has a maximum fractional difference of $0.19\%$. The standard X-PSI resolution waveform exhibits larger variations with fractional differences below $0.13\%$ across most phases, and a maximum fractional difference of $0.76\%$ at the trough.  All these variations, however, would be subsumed within the expected Poisson noise levels for NICER observations. 

The Alberta waveform exhibits generally higher fractional differences with a maximum fractional difference of $0.52\%$ against the ultra X-PSI waveform and $0.47\%$ against the IM waveform, both occurring at the waveform trough. The low resolution X-PSI setting has higher fractional differences compared to the others and has the poorest maximum fractional difference of $2.59\%$ in the lowest count phase bin. Deviations exhibited by both the Alberta and the low resolution X-PSI waveform at the waveform crests are comparable to the Poisson fluctuation in those bins, with the latter far exceeding the noise levels at the troughs.

The patterns in the fractional spectral differences are highly similar to those seen for the Ring-Eq configuration. The differences in values are related to the waveform differences. The only noticeable difference in pattern is the higher number of counts registered by the low X-PSI resolution spectra in the lower channels before the oscillations become dominant, which is a result of over-predicting counts at the waveform crest.

\subsection{Test: Crescent-Polar spot}
\label{sec:pst-polar}

The Crescent-Polar configuration consists of a similarly elongated spot as the Crescent-Eq, but placed such that it encompasses the rotational North Pole, rendering only the very limbs of the spot directly visible at the observer inclination angle as the star rotates. As a result, the pulse profile consists of a relatively low number of counts ($\sim 440$) in the brightest phase-energy bins, and exhibits a prominent pulse centered at around a phase of 0.25 cycles and a fainter pulse after half a cycle. 

We had already noted that a waveform generated by such a shape is relatively sensitive to the different code treatments and resolution settings in Section \ref{sec:pst-eq}. Such sensitivities are heightened in the case of Crescent-Polar as edge-effects dominate (see Appendix \ref{appendix}) the resulting waveform (Figure \ref{fig:pst-polar frac diffs}). The fractional difference is the least for the high resolution X-PSI waveform with values mostly under $0.2\%$ and a maximum of $2.78\%$ at the waveform minima, which occurs in the trough around 0.55 cycles. These differences are within the expected Poisson noise for this waveform configuration assuming NICER observations.

The standard X-PSI resolution, IM and Alberta waveforms show roughly similar levels of fractional differences compared to the ultra X-PSI resolution waveform in most phases beside the minima, typically below $5\%$. Such deviations are comparable to the corresponding Poisson noise level in those bins. However, the maximum fractional differences at the waveform minima for each of these waveforms are $38.05\%$, $23.44\%$, and $30.19\%$  respectively with respect to the X-PSI ultra waveform. The maximum fractional difference between the Alberta and IM codes is $43.45\%$. These values are far beyond the expected Poisson level for NICER in the low count phase bins. The different approach for surface discretisations taken by X-PSI, where the hot spot edges consist of partially emitting cells, also add on to the differences exhibited by the IM and Alberta codes as compared against the X-PSI ultra waveform.

The low X-PSI resolution waveform differs the most from the ultra resolution waveform, with large fractional differences of $>10\%$ occurring at both troughs, and a maximum fractional difference of $66\%$ (see top-middle panel of the plot on the left in Figure \ref{fig:pst-polar frac diffs}). Even at phase bins with relatively high count numbers, the low resolution X-PSI waveforms therefore show deviations far greater than the Poisson noise levels. The main culprit behind the large discrepancy between the low and ultra resolution X-PSI settings for this configuration is the number of elements used to construct the hot region cell mesh.

The generally high fractional differences imply the need for more refined discretisations if inferred shapes involve such elongated structures, especially if only the very edges are directly visible, resulting in fewer registered counts.

\begin{figure*}
    \centering
    \includegraphics[width=1\linewidth]{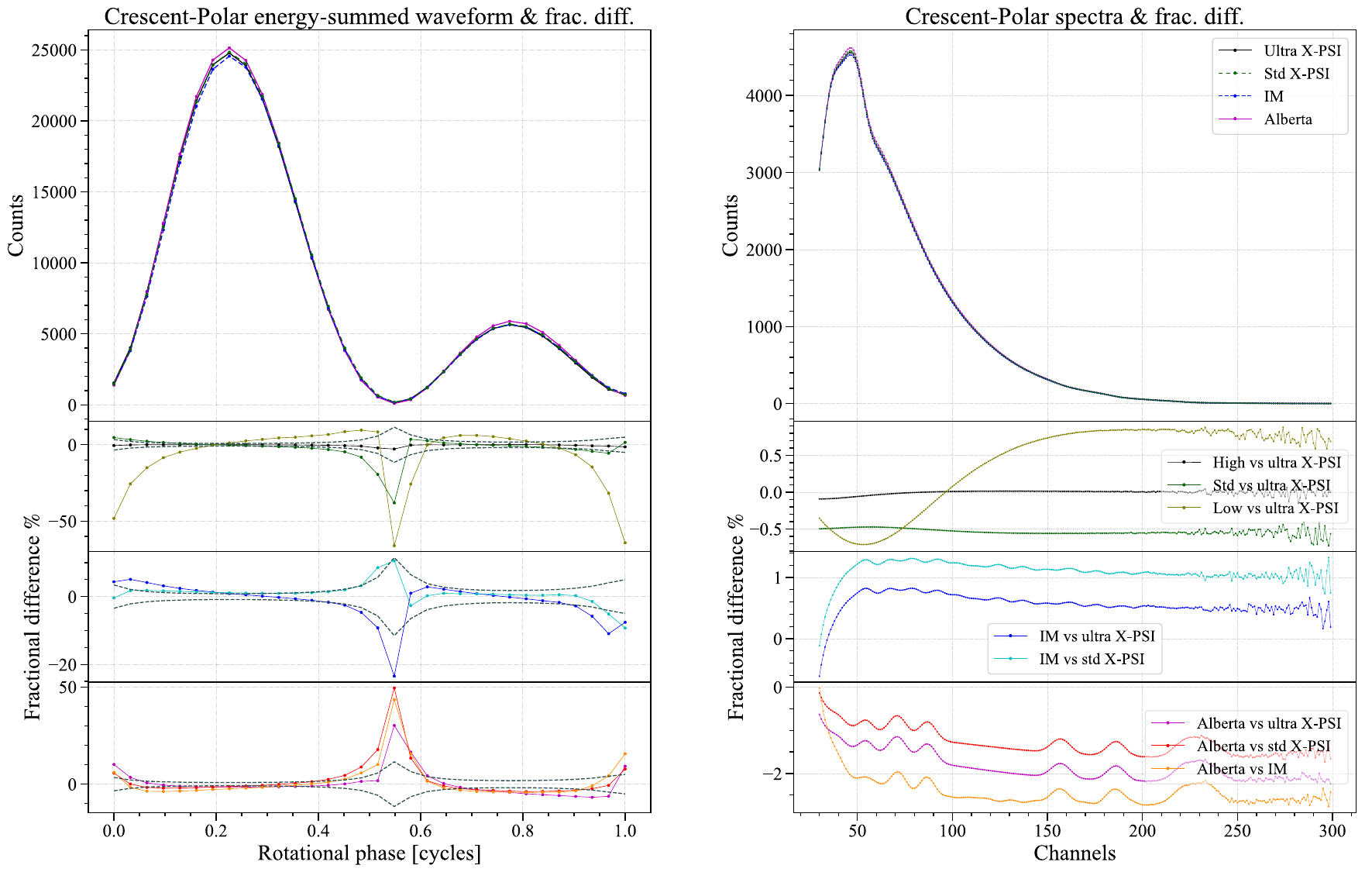}
    \caption{Energy-summed waveforms (left) and phase-summed spectra (right) generated by the different codes for the Crescent-Polar configuration (top) and the corresponding fractional differences.}
    \label{fig:pst-polar frac diffs}
\end{figure*}

Looking at the fractional spectral differences, we note some differences in patterns from the previous waveforms. The standard X-PSI consistently under-predicts the number of counts per channel and exhibits higher differences against the ultra resolution spectra, compared to the corresponding differences between the high and ultra resolution spectra. The low resolution spectra behave quite differently, either under- or over-predicting counts. The oscillations seen previously in the fractional spectral differences are not as prominent in comparison and mostly affect the highest channels which consist of a very few counts.

The IM spectra still retain the energy interpolation related oscillations when compared against the standard and ultra X-PSI spectra, although this time the general amplitude pattern appears slightly different and there is a noticeable tail in channels 30--40. The Alberta spectra seem to retain more of the previously seen amplitude pattern in its oscillations. These spectral differences are a consequence of the hot spot edge-effects.

\subsection{Test: Bithermal spot} 
\label{sec:pdt}

The Bithermal spot waveforms have only been tested between the X-PSI and IM codes due to their ability to handle multiple-imaging. Being placed at the pole, this configuration yields a pulse profile with low modulation, much like the Ring-Polar waveform (Figure \ref{fig:pulse profiles}). Being placed at the rotational north pole, this spot would not have been directly visible from the defined observer inclination angle in the absence of gravitational lensing. We only register a low number of counts ($\sim 230$ counts in the brightest phase-energy bins), which arrive at the telescope only as a result of light-bending (due to the compact nature of the model star) and relatively horizontal emission with respect to the surface. 

\begin{figure*}
    \centering
    \includegraphics[width=1\linewidth]{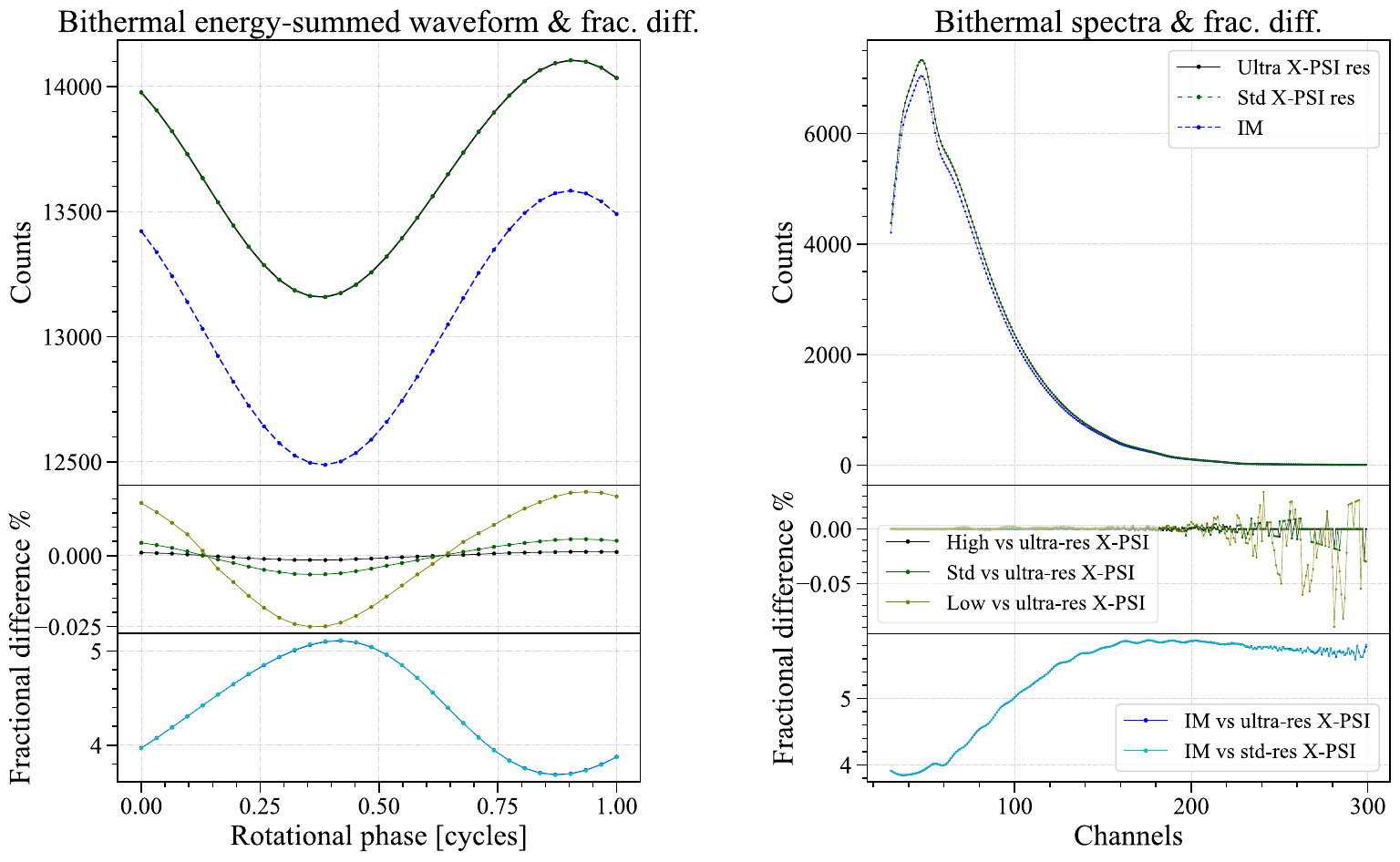}
    \caption{Energy-summed waveforms (left) and phase-summed spectra (right) generated by the different codes for the Bithermal spot configuration (top) and the corresponding fractional differences. For clarity, the Poisson bands (at about 1.2\%) are not shown in this figure.}
    \label{fig:pdt-polar frac diffs}
\end{figure*}

The waveform generated by the IM code is noticeably different to those generated by X-PSI (Figure \ref{fig:pdt-polar frac diffs}). The mean fractional difference between the IM and the X-PSI generated waveforms is $4.4\%$, and the maximum fractional difference is $5.11\%$ at the waveform trough. This is much higher than the expected Poisson fluctuation for NICER. The maximum Poisson noise is at a $1.23\%$ level for the X-PSI ultra waveform scaled to $10^6$ counts.

Unlike X-PSI, where ray-tracing is performed on the fly, the IM waveform is generated from pre-computed lookup tables where the smallest ratio of $c^2R_{\rm eq}/GM = 3.1$. The ratio for this configuration is 3.038. The extrapolation therefore takes place at a regime where waveforms change rapidly in relation to this ratio. Coupled with differences in surface discretisation treatments as applied to an overlapping spot configuration, and the spot placement requiring high deflection angles to be visible, we end up with such levels of discrepancies between the two codes. For less extreme geometries explored in \citetalias{Bogdanov21} where the spot is prominently visible, the agreement is better, typically around $2\%$, between the IM and X-PSI code, for the same high compactness scenario.

The different X-PSI resolutions are quite consistent with each other for this configuration. The low resolution setting waveform shows the largest fractional difference, with a maximum fractional difference of $0.03\%$ at the waveform trough. 

The fractional spectral difference is nominal between the different X-PSI resolutions, with some energy resolution dependent oscillations towards the higher channels as seen previously. The IM spectrum shows much larger differences when compared to any of the X-PSI spectra. We can once again see the energy interpolation dependent oscillations, but these effects are dwarfed by the differences arising from the varying ray-tracing calculations for such an extreme model configuration.

\section{Quantitative Differences Between Waveforms}
\label{sec:differences}

Parameter inference using NICER data follows the standard Bayesian procedure: for a given set of model parameters, we compute the likelihood of the NICER data given the model waveform and multiply that likelihood by the prior probability density at those parameters to obtain the unnormalized posterior probability density.  If the waveforms are inaccurate then this can introduce errors in our inference.  However, if the errors are smaller than the Poisson fluctuations in the data then the errors are subdominant in the overall process of inference.

\begin{table*}
\renewcommand{\arraystretch}{1.5}
    \centering
    \caption{$\chi^2$ versus X-PSI Ultra}\begin{tabular}{l c c c c c} \hline \hline
        \textbf{Geometry}& \textbf{X-PSI High}& 
        \textbf{X-PSI Std} & 
        \textbf{X-PSI Low}& 
        \textbf{IM}& 
        \textbf{Alberta}\\ \hline
                 \textbf{Phase-channel $\chi^2$} & & & & & 
                 \\
         Ring-Eq & 0.002& 0.035& 1.085& 0.742&44.243\\
         Ring-Polar& 0.0002 & 0.0009 & 0.008& 0.224&25.24\\
         Crescent-Eq & 0.178& 3.134& 37.905& 0.718&25.913\\
         Crescent-Polar& 5.083&414.528& 5600.30& 480.426&726.932\\
         Bithermal& 0.0003& 0.002& 0.03& 2088.61\\
         \ \ \ \ $+$normalization & & & & 67.949\\
         \hline
                \textbf{Bolometric $\chi^2$} & & & & & \\
         Ring-Eq & 0.002& 0.034& 1.036& 0.594&33.668\\
         Ring-Polar & $1.14\times 10^{-7}$& $2.72\times 10^{-6}$& $7.62\times 10^{-6}$& 0.152&16.977\\
         Crescent-Eq& 0.176& 3.094& 37.38& 0.646 &16.102\\ 
         Crescent-Polar& 4.615& 402.052& 5508.08& 448.527 &695.77\\ 
         Bithermal& 0.0001& 0.002& 0.027& 2046.06&\\
         \ \ \ \ $+$normalization & & & & 27.271\\
         [1.0ex] \hline
    \end{tabular}
    \tablecomments{$\chi^2$ comparisons of different waveforms with the waveform produced in the X-PSI ultra run. The top block shows comparisons of the full set of phase-channel bins (32 rotational phases, 270 channels) and the bottom block shows comparisons of the bolometric $\chi^2$ (i.e., summed over channels).  For each spot geometry we show the $\chi^2$ for (left to right) the X-PSI high, X-PSI std, X-PSI low, IM, and Alberta waveforms.  There is excellent agreement between the two production codes (X-PSI std and IM) for the Ring-Eq, Ring-Polar, and Crescent-Eq configurations.  The large difference in the Bithermal configuration is due to the high compactness being outside the table boundaries for the IM code, but most of the difference disappears when an energy-independent normalization is added, equivalent to a small change in the assumed distance to the source or effective area of the instrument.  The large difference in the Crescent-Polar configurations is attributable to different assumptions about the spot-boundary cells between the X-PSI and the IM and Alberta codes.  See text for details.}
    \label{tab:chisq}
\end{table*}

Our analysis uses Poisson likelihoods, given the expected statistical independence of the counts recorded with NICER.  The number of counts in NICER data is large enough that Wilks' theorem \citep{Wilks1938} applies and the difference $\Delta\ln{\cal L}$ between log likelihoods is related to the difference $\Delta\chi^2$ between chi squareds via $\Delta\ln{\cal L}\approx -\frac{1}{2}\Delta\chi^2$.  We can therefore use $\Delta\chi^2$ as a quantitative measure of the difference between waveforms.  Because
\begin{equation}
\chi^2=\sum_i\frac{(m_i-d_i)^2}{m_i}
\label{eq:chisq}
\end{equation}
(where $m_i$ is the expected model counts in data bin $i$ and $d_i$ is the observed number of counts in bin $i$) is a nonlinear measure, the phase-channel $\Delta\chi^2$ (for which we use all 32 rotational phase bins and all 270 channels) and the bolometric $\Delta\chi^2$ (for which we sum over all channels and use their sum in each of the 32 phase bins) can reveal different waveform discrepancies.  We therefore use both.  

The acceptable threshold value for $\Delta\chi^2$ between model waveforms is somewhat subjective.  For $N$ parameters, where in our models $N\sim 10-20$, $\approx 50$\% of the probability is contained within $\Delta\chi^2=N$ of the minimum $\chi^2$, so if $\Delta\chi^2>N$ between models then model inaccuracies could alter inferences significantly.  

In our comparisons we arbitrarily use the X-PSI ultra waveform as the ``data" and other waveforms as the ``models".  We then have three practical considerations for our $\Delta\chi^2$ test.  The first has to do with the number of counts: from Equation~\ref{eq:chisq} it is clear that the expected $\Delta\chi^2$ is proportional to the number of counts.  We therefore normalize each waveform by a factor such that the total number of counts in the X-PSI ultra waveform equals $10^6$; because the same factor (and thus effectively the same exposure time) is used for all waveforms, the other waveforms generally have slightly more or less than $10^6$ total counts.  The second consideration relates to the overall, energy-independent, normalization of the waveform.  There are two parameters commonly used in NICER inferences (the distance to the neutron star and the effective area normalization for NICER) which have the same effect: they simply scale the entire waveform up or down.  If the primary difference between waveforms is an overall normalization, then this will be absorbed in such factors and will not affect the inferred posterior for the other parameters.  The third consideration is that, in addition to the NICER counts from hot spots, there are counts from other sources, e.g., a particle background, optical loading, or background sources.  These added counts increase the level of Poisson fluctuations and thus diminish the impact of waveform inaccuracy.  However, to be conservative, we do not include backgrounds in our waveform comparisons.

Table~\ref{tab:chisq} shows the phase-channel and bolometric $\Delta\chi^2$ values for our waveforms. Because we are using the waveforms without Poisson sampling, $\chi^2=0$ is possible and is nearly achieved in several cases.  Based on these measures, the production codes (X-PSI std and IM) agree well with each other for the Ring-Eq, Ring-Polar, and Crescent-Eq configurations.  The Bithermal configuration has a compactness above that tabulated in the IM code, but when a free energy-dependent normalization is included the agreement improves substantially.  The Crescent-Polar configuration is sensitive to the treatment of the cells at the boundary of the spot, which is different in the X-PSI code versus the IM or Alberta codes.  We conclude that for typically obtained spot configurations in NICER analyses, when the codes make the same assumptions their waveforms are close enough that in a practical sense they do not have an adverse effect on parameter inference.

\section{Discussion and Conclusions}
\label{sec:discussion}

In this work, we have tested five extreme configurations that lie in the model spaces of the X-PSI, IM and Alberta ray-tracing codes. These tests probed the effects on the generated pulse profiles by the different exact treatments of the various codes regarding spot overlap, elongated shapes, and multiple-imaging for compact stars. 

In \citetalias{Bogdanov19b}, the fractional differences exhibited by the bolometric waveforms generated by the different codes for simple emission geometries were typically $\lesssim 0.1$\% (well below the expected NICER Poisson noise level), except when the flux contributions were very low at the waveform minima. The fractional differences for high-compactness tests in \citetalias{Bogdanov21}, again consisting of simple emission geometries, were slightly higher. 

The tests in this paper exhibit varying levels of fractional differences depending on the individual configurations considered. We checked whether the fractional differences for our tests exceeded the corresponding expected NICER Poisson noise levels in the respective phase bins. We additionally tested whether the $\Delta \chi^{2}$ (full phase-channel resolved and bolometric), which stands as a proxy for the difference in log likelihoods, between the different waveforms and the X-PSI ultra waveform is high enough to affect inferences, particularly when using the production codes, X-PSI std and IM.

In general, the overlap treatment between the different codes yields consistent results, even when considering tiny emission regions and pole coverage, as demonstrated by the fractional differences between the different waveforms and spectra for the Ring-Eq and Ring-Polar tests. For both of these configurations, the differences between X-PSI and IM waveforms are well below the expected Poisson noise across all phases. Consequently, the $\Delta \chi^{2}$ values also show a high degree of agreement between the production codes, and with X-PSI ultra. The differences exhibited by the Alberta waveform in comparison to the others are either below the noise level or equivalent to it in some of the phase bins.

The effects of the varying surface discretisation treatments begin to manifest in the pulse profiles generated when we consider highly elongated and pointed shapes. The Crescent-Eq and Crescent-Polar cases show larger differences compared to the Ring-Eq and Ring-Polar cases. This is primarily due to the different way the spot edges are handled by X-PSI compared to that of IM and Alberta codes. The contribution of the boundary cells plays a more important role for such elongated shapes, as they form a significant fraction of the spot's surface area. 

For the Crescent-Eq configuration, where the whole spot goes in and out of view, the fractional differences between the X-PSI (except the low resolution version) and IM waveforms are much less than the Poisson uncertainty over all phases. Therefore, the $\Delta \chi^{2}$ value again shows that the production codes agree well with each other, and with X-PSI ultra. The low resolution X-PSI and Alberta waveforms exhibit differences either lower than the Poisson noise level, or equivalent to it in some phase bins.

Depending on the viewing angle for such elongated geometries, the differences can be more pronounced, as evidenced by the Crescent-Polar case, where the bulk of the emission from within the spot is not visible and we mostly see only the limbs of the spot. Here, the resolution of the spot cell-mesh discretisation plays a more significant role. Only the high resolution X-PSI waveform exhibits fractional differences lower than the corresponding Poisson noise level across all phases when compared against the ultra resolution X-PSI waveform. All other waveforms, when compared to X-PSI ultra, mostly show differences equivalent to or larger than the noise level. The differences are especially stark for the low resolution X-PSI waveform where the spot shape is very crudely approximated. The $\Delta \chi^{2}$ values indicate that, for such scenarios, the treatment of the mesh could unpredictably affect the likelihood surface during parameter inference and that this disagreement cannot be mitigated even when additionally considering energy-independent normalization of the waveform (meant to mimic the effect of instrument effective area normalization during the inference process).

The Bithermal configuration was only tested between X-PSI and IM codes, since Alberta does not incorporate multiple-imaging for its ray-tracing, which plays a crucial role for high compactness scenarios as is the case with this test. We find high fractional differences between IM and X-PSI waveforms and spectra, with the waveform differences being higher than the expected noise levels across all phases. This is a consequence of the compactness for this model exceeding the tabulated values for the IM code. However, the $\Delta \chi^{2}$ metric indicates that these differences can largely be compensated by the instrument normalization factor during inference.

For all of the configurations tested, the spectral fractional differences exhibit an oscillatory behaviour that is linked to the different interpolation schemes adopted by the different codes, becoming more pronounced in the higher channels containing low numbers of counts. In the case of Crescent-Polar, the spot boundary treatments also additionally contributes with effects that are entangled with its impact on the waveforms.

Our tests thus indicate that, even for more complex surface emission scenarios, the pulse profiles generated by the three ray-tracing codes are mostly in agreement, and that parameter inferences using NICER data by the production codes, X-PSI std and IM, should remain largely unaffected by the differing details of their numerical treatments. However, there are certain scenarios where users of these codes should exercise more caution. If the model spaces considered allow for sampling of large elongated structures, and the posteriors indicate preference for such geometries especially at viewing angles where edge effects dominate, then it is advisable to validate the inferred results using high resolution settings, particularly those concerning the surface cell-mesh discretisation. Alternatively, if such configurations can be deemed unrealistic based on detailed pulsar magnetospheric simulations, or through joint fits using data collected at other wavelengths, the model prior spaces can be constrained to exclude such scenarios. The latter approach might be especially worthy of consideration keeping computational expense incurred by high resolution runs in mind. 

All three codes tested in this paper currently employ the OS approximation for the spacetime embedding. The effect of the choice of metric has been tested in a number of papers by embedding an oblate shape in different metric approximations and comparing the resulting waveforms with waveforms constructed with an exact numerically generated metric. For the relatively slow spin rates of the primary pulsar populations targeted by NICER ($<400$ Hz), the errors with respect to the numerical solution are not significant \citep{Cadeau07_thesis, Bogdanov19b, Silva21}. For faster spin rates expected of accreting/bursting neutron stars, the shape of the surface and the choice of the metric can introduce errors depending on the geometry \citep{Cadeau07, Pihajoki2018, Silva21}. Therefore, an interesting and useful follow-up would be to perform systematic tests of the OS approximation for spin rates in the range of $200 - 600$ Hz using extreme geometric configurations like the ones in this paper, especially obscure and elongated structures such as the Crescent-Polar case.


\begin{acknowledgments}
D.C., A.L.W., T.S. and S.V. acknowledge support from ERC Consolidator Grant No.~865768 AEONS (PI: Watts).  M.C.M. and A.J.D. acknowledge funding from NASA ADAP grants 80NSSC20K0288 and 80NSSC21K0649.  S.M. acknowledges support from NSERC Discovery Grant RGPIN-2019-06077.  S.B.~acknowledges funding from NASA grants 80NSSC20K0275 and 80NSSC22K0728. S.G. acknowledges the support of the CNES. M.T.W. acknowleges support by the NASA Astrophysical Explorers Program.We acknowledge extensive use of NASA’s Astrophysics Data System (ADS) Bibliographic Services and the ArXiv.
\end{acknowledgments}

\software{\texttt{X-PSI} \citep{Riley23}, \texttt{Jupyter} \citep{2007CSE.....9c..21P, kluyver2016jupyter}, \texttt{matplotlib} \citep{Hunter:2007}, \texttt{numpy} \citep{numpy}, \texttt{python} \citep{python} and \texttt{Cython} \citep{cython:2011}.}
Software citation information aggregated using \texttt{\href{https://www.tomwagg.com/software-citation-station/}{The Software Citation Station}} \citep{software-citation-station-paper, software-citation-station-zenodo}.

\bibliographystyle{aasjournal}
\bibliography{Cross-tests}

\clearpage 
\appendix
\section{Hot spot cell mesh prescriptions and resolution}\label{appendix}

In Section \ref{sec:pst-polar}, we discussed how the resolution of the cell meshes describing the shape of the hotspot on the stellar surface affects the consequent pulse profile. Additionally, the differing mesh prescriptions between the three codes further contribute to the discrepancies between them. Here we outline the main differences between the different codes in this regard. For a more detailed description, we refer the reader to Appendix B of \citetalias{Bogdanov19b}.

\begin{figure*}[h]
    \centering
    \includegraphics[width=\linewidth]{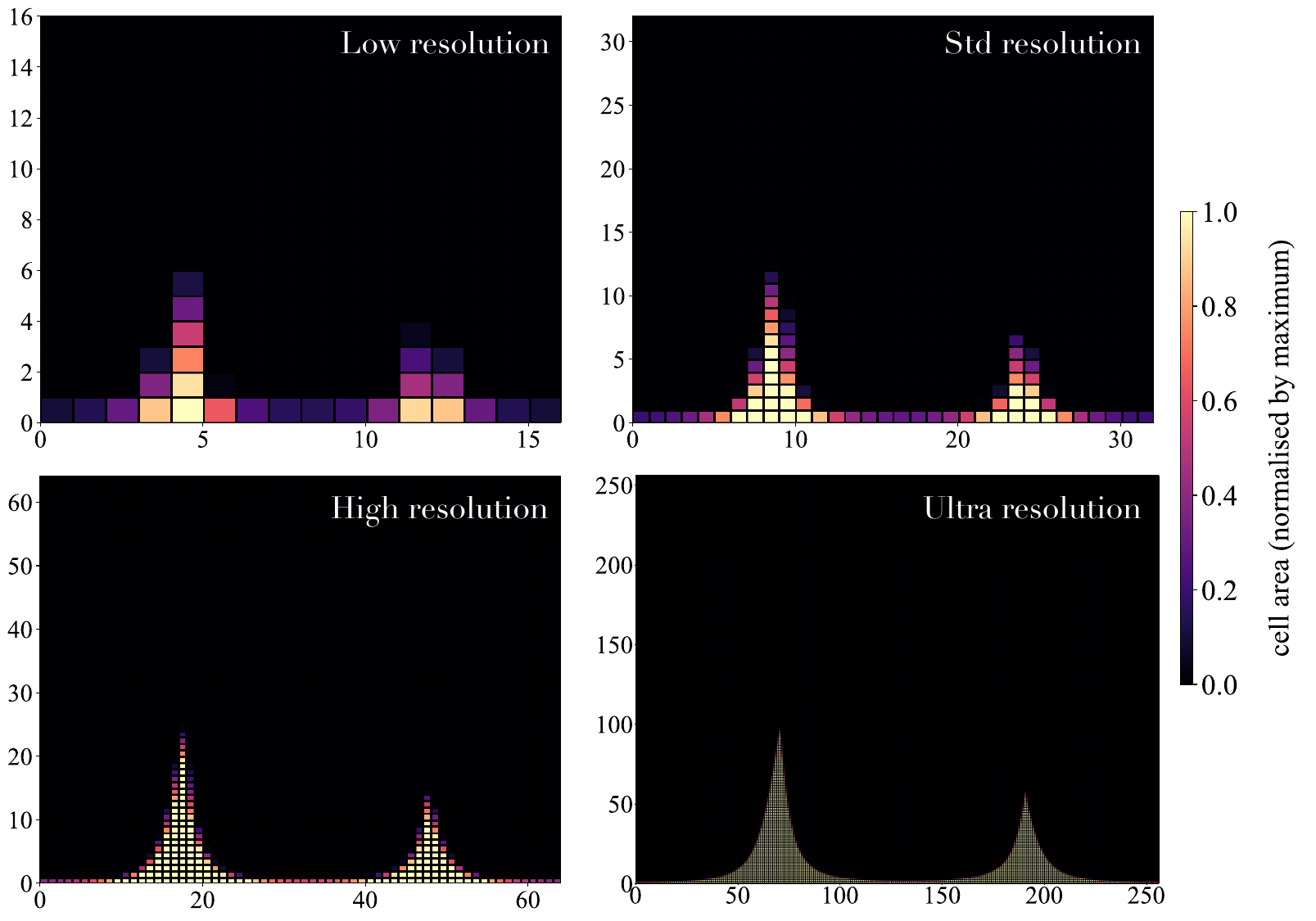}
    \caption{Spot cell mesh discretisation for the Crescent-Polar configuration for the different X-PSI resolution settings. The axes represent the total number of cell mesh elements considered for construction of the hotspot along the azimuth (x-axis) and colatitude (y-axis). The boundary in the azimuthal direction is cyclical because the spot region extends over one of the poles, while the lower boundary in colatitude encompasses a coordinate singularity. The colours depict the proper area occupied by the hot region within each mesh element, which is used to weight the contribution of that cell.As a result, the cells at the boundary of the spot only emit partially, as compared to the cells within. Since the visible area is rather small and elongated for this configuration, edge effects become significant, leading to a poor approximation when using a coarse mesh resolution.}
    \label{fig:cellmesh}
\end{figure*}

In X-PSI, the hot spot cell mesh is discretized into a regular mesh of points in colatitude and azimuth (about the stellar spin axis). The number of mesh elements are user-defined by setting the \texttt{sqrt\_num\_cells} (See table \ref{tab:resolutions}). Figure \ref{fig:cellmesh} demonstrates the X-PSI mesh treatment and the effect of the various mesh resolution settings tested in this work as applied to the Crescent-Polar configuration. A unique aspect of X-PSI is that the boundary cell meshes only emit partially, as a function of the area of the mesh element that's covered by the hot spot. This approach allows for an exact areal determination of the spot. This method of using partially emitting cell meshes is only used in X-PSI, and not in the IM and Alberta codes.
 
In the IM code, each spot is divided into slices of equally-spaced colatitude (with 200 slices as the default).  At each colatitude, the slice is represented as equally-spaced segments of longitude (with 100 segments as the default).  Each colatitude/longitude pair is then treated as a Cartesian pixel for the purpose of discretizing the spot.

The Alberta code does a similar discretization of the spot, as the IM code, based on its colatitude and longitude. The number of segments is chosen by creating a sequence of spots with increasing numbers of discrete segments until the signal converges.

\end{document}